\documentclass[11pt,twoside]{article}
\usepackage{asp2004}
\usepackage{psfig}
\usepackage{epsf}
\usepackage{graphics}
\usepackage{lscape}
\begin{document}
\title{Outbursts of WZ Sagittae}

\author{O. M. Matthews$^1$, M. R. Truss $^2$, G. A. Wynn$^1$, R Speith$^3$}
\affil{
$^1$ Department of Physics \& Astronomy, University of Leicester, University Road, Leicester, LE1 7RH, United Kingdom\\
$^2$ School of Physics \& Astronomy, University of St Andrews, North Haugh, St Andrews, Fife, KY16 9SS, United Kingdom\\
$^3$ Institut f{\"u}r Astronomie und Astrophysik, Universit{\"a}t T{\"u}bingen, Auf der Morgenstelle 10C, D-72076 T{\"u}bingen, Germany}

%\author{O. M. Matthews}
%\affil{Department of Physics \& Astronomy, University of Leicester, University Road, Leicester, LE1 7RH, United Kingdom}
%\author{M. R. Truss}
%\affil{School of Physics \& Astronomy, University of St Andrews, North Haugh, St Andrews, Fife, KY16 9SS, United Kingdom}
%\author{G. A. Wynn}
%\affil{Department of Physics \& Astronomy, University of Leicester, University Road, Leicester, LE1 7RH, United Kingdom}
%\author{R. Speith}
%\affil{Institut f{\"u}r Astronomie und Astrophysik, Universit{\"a}t T{\"u}bingen, Auf der Morgenstelle 10C, D-72076 T{\"u}bingen, Germany}

\begin{abstract}
The outbursts of the dwarf nova WZ Sagittae occur on timescales which are much longer than expected from the standard thermal viscous disc instability model, which correctly reproduces the outburst behaviour of many other dwarf novae. In addition short `echo' outbursts have sometimes been observed after the main outburst. The increased timescales could be explained by a magnetic propeller, anchored to the primary star, which clears the inner disc. We argue that the propeller, combined with the resonances of the binary orbit, may also be the cause of the echo outbursts. We present numerical models which reproduce the echo outbursts.  
\end{abstract}

\section{Introduction}
WZ Sagittae is a member of the SU Ursae Majoris class of dwarf novae. These are short period systems which exhibit long superoutbursts in addition to ordinary outbursts. During these superoutbursts periodic modulations of the optical light curve, known as superhumps are observed. These features have a frequency which is slightly longer than that of the binary orbit, and which is usually attributed to a precessing disc. Superoutbursts in the SU UMa systems typically occur with recurrence times of weeks or months.

Stars from the subclass to which WZ Sge belongs are sometimes known as TOADs or Tremendous Outburst Amplitude Dwarf novae. These systems have very long recurrence times and correspondingly bright superoutbursts. WZ Sge itself showed a 33 year recurrence time, until this pattern was broken by the 2001 outburst which occurred after only 23 years. WZ Sge, the brightest of all dwarf novae, is also notable for its extreme mass ratio. According to \citet{ski02} $M_{2}=0.045  \pm 0.003 \ M_{\sun}$ and $M_{1} = 0.74 \pm 0.07 \ M_{\sun}$. It has a binary period of $P_{\rm{orb}} = 81.6 \ {\rm{min}}$ \citep{krz62} and a primary spin period of $P_{\rm{spin}} = 27.87 \ {\rm{s}}$. In this outburst \citep[e.g][]{pat02} superhumps were observed, but prior to these a modulation at the frequency of the binary orbit was observed by \citet{pat02} as discussed by e.g. \citet{kun04}. In addition, the main outburst was followed by twelve echo outbursts. Echoes have also been observed in the similar system EG Cancri \citep{osa01}.

Numerical models incorporating the thermal viscous disc instability model (DIM) due to \citet{hos79}, have reproduced the outburst behaviour of many dwarf noave. These models have not however, without using extremely low viscosities, been able to explain the long recurrence time of WZ Sge type systems. \citet{osa89} extended the DIM to binaries with small mass ratios when \citet{whi88} showed that discs in such systems could become tidally unstable. Osaki proposed that, during the normal outburst cycle, material is allowed to build up in the outer disc. When the disc becomes sufficiently large that it may access the 3:1 resonance of the binary orbit, it will become eccentric, enhancing dissipation and the loss of angular momentum, leading to a superoutburst. Simulations \citep[e.g.][]{tru01} supported this theory.

The extended recurrence times of the WZ Sge type systems may be explained by the use of a magnetic propeller to evacuate the inner regions of the disc. This mechanism, similar to that proposed by \citet{mat04a} for the FU Orionis stars, is detailed in the forthcoming paper by \citet{wyn04}. The magnetic propeller is a mechanism by which the inner regions of an accretion disc can be evacuated. The co-rotation radius $R_{\rm{co}}$ is the radius at which a Keplerian orbit has the same angular frequency as that of the star's spin. If the primary magnetic field rotates as a solid body then this radius can be said to divide the magnetic accretion and magnetic propeller regimes. Particles at radii smaller than the co-rotation radius lose angular momentum as a result of their interactions with the stellar magnetic field, and fall rapidly onto the star. This magnetically enhanced accretion occurs because the particles are moving more rapidly than the magnetic field. Elements of plasma which exist outside co-rotation are, conversely, accelerated by the magnetic field, and gain angular momentum at the star's expense. This is the magnetic propeller regime. An example of a strong magnetic propeller is that of AE Aquarii \citep{wyn97}, in which accretion is completely prevented by the magnetic field. What is proposed here is a far weaker propeller, driven by a rapidly spinning, moderately magnetic white dwarf. Simulations using weaker fields have been shown to warp accretion discs \citep{osu04}. \citet{las99} also used a magnetically truncated disc in their model of WZ Sge. 

The cause of the echo outbursts is as yet uncertain. \citet{osa01} suggested that the mass which accumulates near the 3:1 resonance in the disc is released gradually after the main outburst, leading to echo outbursts. In this paper we model a full outburst of WZ Sge using smoothed particle hydrodynamics (SPH). The rapidly spinning white dwarf creates a magnetic propeller which, in addition to evacuating the inner disc, aids the creation of a reservoir of material in the outer disc. Echo outbursts arise naturally from this configuration.

\section{Numerical Method}
The simulations in this paper were performed using a fully three dimensional SPH code with viscosity switching and a magnetic acceleration as described below. The specific code is fully described by \citet{tru00} whilst \cite{mon92} provides a general review of SPH. SPH is a Lagrangian technique for the modelling of fluid dynamics. It works by discretising the fluid into a series of particles, normally with equal mass, which represent fluid elements. These particles carry values for fluid properties such as density. The properties of the fluid may then be interpolated for any point in space by taking a weighted average of the properties of nearby particles. This is done using a smoothing kernel, which usually takes the form of a cubic spline function. 
%For example a function of position  such as $f\left(\mbox{\boldmath{$r$}}\right)$ is interpolated from nearby particles using the expression
%\begin{equation}
%\label{eqn:interp}
%f\left(\mbox{\boldmath{$r$}}\right) = \sum_{\rm{i=1}}^{\rm{N}} \frac{m_{\rm{i}}}{\rho_{\rm{i}}} f_{\rm{i}} W \left( {\mbox{\boldmath{$r$}}} - {\mbox{\boldmath{$r$}}_{\rm{i}}},h \right)
%\; ,
%\end{equation}
%where the sum is over all $N$ particles and $m_{i}$, $\rho_{i}$, $f_{i}$ and $\mbox{\boldmath{$r$}}_{\rm{i}}$ are the mass, density, value of function f and position of the $i$th particle respectively. The smoothing kernel is represented by $W$ and the smoothing length, or characteristic length of this kernel, is denoted by $h$. This formulation works because $\rho_{\rm{i}}/m_{\rm{i}}$ is a volume element. 

The viscosity in the disc is an artificial viscosity which is introduced into the velocity and energy equations of the SPH formulation. The extra term is composed of shear and bulk viscosity elements. However, providing that the local velocity divergence is negligible, it can be shown \citep{mur96} that the viscosity acts in a manner very similar to the \citet{sha73} $\alpha$ shear viscosity. The SPH viscosity parameter $\zeta$, which is proportional to $\alpha$, is switched between two values, mimicking the behaviour of the DIM. Switching occurs at two trigger surface densities. Any annulus of the disc which exceeds $\Sigma^{\rm{crit}}_{\rm{high}}$ will begin to have its $\zeta$ gradually altered to the high state value. Conversely, an annulus which falls below $\Sigma^{\rm{crit}}_{\rm{low}}$ will have its $\zeta$ value slowly reduced to the lower, quiescent value. This is discussed by \citet{tru00}. The positions of the triggers, which are roughly proportional to radius, are dervied by \citet{can88}.

\citet{den90} expresses the magnetic tension acceleration as
\begin{equation}\label{neweq2}
a_{\rm{mag}} \sim \frac{1}{\rho r_{\rm{c}}} \left( \frac{B_{{z}} B_{{\phi}}}{4 \pi} \right)
\; ,
\end{equation} 
where $\rho$ is density and $r_{\rm{c}}$ represents the radius of curvature of the magnetic field lines. The azimuthal and vertical components of the magnetic field are represented by $B_{\rm {\phi}}$ and $B_{\rm {z}}$ respectively. We will make the approximation that the radius of curvature of the field lines is of the order of the disc scale-height thus, $r_{\rm{c}} \sim H$
\citep[e.g.][]{pea97}. The ratio of vertical and azimuthal field strengths is
related to the shear between the disc and the magnetic field. If it is
assumed that the field rotates rigidly with the star then this ratio can be
expressed in the form \cite[e.g.][]{liv92} 
\begin{equation}\label{neweq3}
\frac{B_{{\phi}}}{B_{{z}}} \sim - \frac{\left(\Omega_{\rm{k}} - \Omega_{\rm{\star}} \right)}{\Omega_{\rm{k}}}
\; ,
\end{equation}
where $\Omega_{\star}$ indicates the angular frequency of the accreting star while $\Omega_{\rm{k}}$ represents the Keplerian angular frequency at a given angle. Equations (\ref{neweq2}) and (\ref{neweq3}) are combined. Additionally it is assumed that the field is dipolar so that the magnetic field can be related to the magnetic moment and distance from the star by the expression $B_z \sim \left|\mbox{\boldmath{$\mu$}}\right|R_{\star}^{-3}$ where $R_{\star}$ represents the radius of the star. It is also assumed that surface density can be approximated by $\Sigma \sim \rho H$, where $H$ is the scale height of the accretion disc. This allows us to make the following approximation
\begin{equation}\label{neweq5}
a_{\rm{mag}} \sim \frac{B_{\rm{z}}^{2}}{4 \pi \rho H} \frac{\left(\Omega_{\rm{k}} - \Omega_{\rm{\star}} \right)}{\Omega_{\rm{k}}} = - \frac{\mbox{\boldmath{${\mu}$}}^{2} R^{-6}}{4 \pi \Sigma} \frac{\left(\mbox{\boldmath{${v}$}}_{\rm{k}} - \mbox{\boldmath{${v}$}}_{\rm{\star}} \right)}{\mbox{\boldmath{${v}$}}_{\rm{k}}}  \sim  - k_{\rm{m}} R^{-6} \frac{\left(\mbox{\boldmath{${v}$}}_{\rm{k}} - \mbox{\boldmath{${v}$}}_{\rm{\star}} \right)}{\mbox{\boldmath{${v}$}}_{\rm{k}}} 
\; . 
\end{equation}
In this fomulation the acceleration falls to zero when $\mbox{{\boldmath{$v$}}}_{\rm{k}} = {\mbox{\boldmath{$v$}}}_{\rm{\star}}$ at co-rotation as expected. It may be useful to compare the value constant $k_{m}$ with the $\beta$ used in \citet{mat04b}. It can be shown that
\begin{equation}
\label{neweq13}
k_{\rm{m}} = - \frac{\beta \sqrt{G M_{\star}}}{2 \Sigma}
\ {\rm{where}} \
\label{neweq14}
\beta = \frac{\mbox{\boldmath{${\mu}$}}^{2}}{2 \pi \sqrt{G M_{\star}}} 
\; ,
\end{equation}
where $G$ is the universal constant of gravitation and $M_{\star}$ is the mass of the star. 

\begin{figure*}[t]
\begin{center}
\resizebox{120mm}{40.0mm}{
\mbox{
\includegraphics{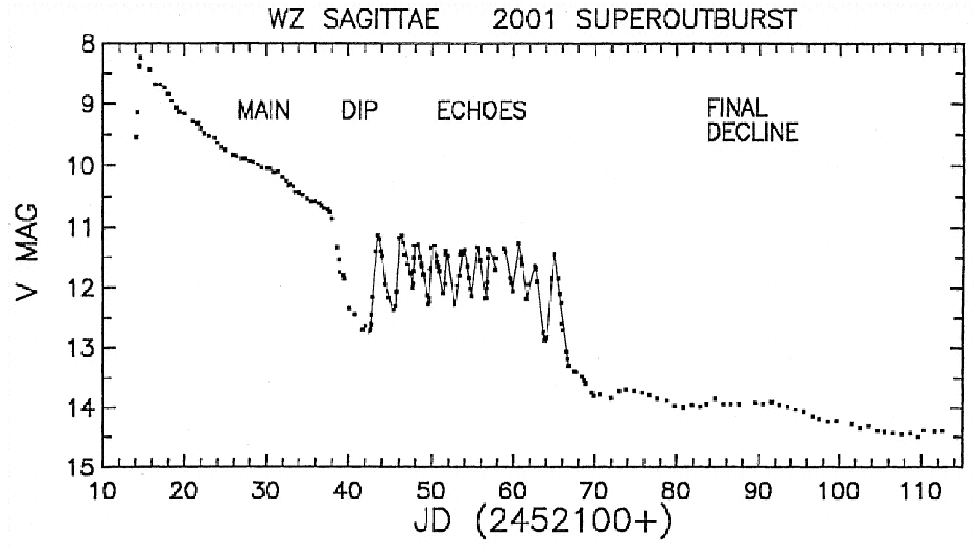}}}
\resizebox{140mm}{40.0mm}{
\mbox{
\includegraphics{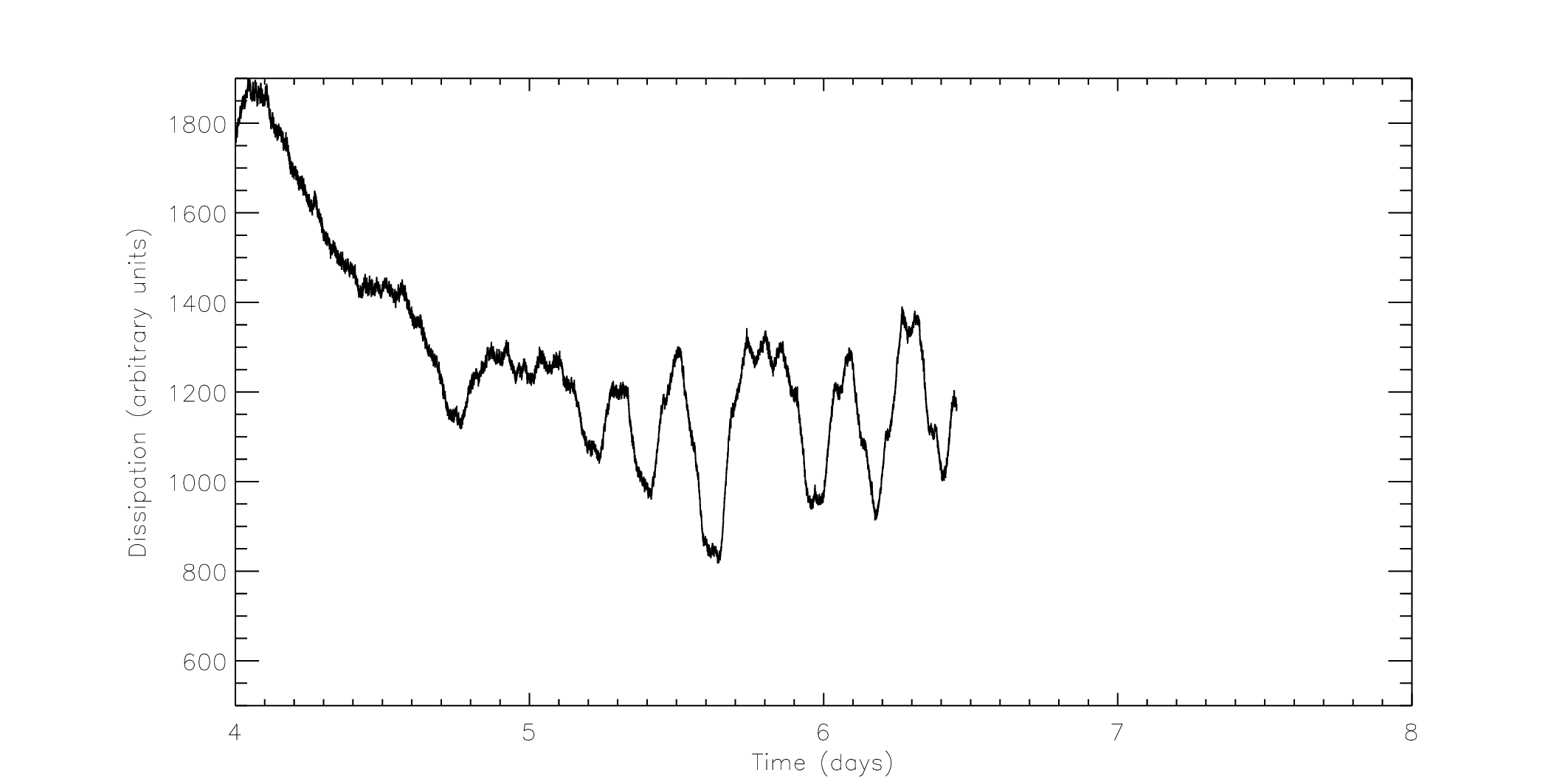}}}
\end{center}
\caption{{\it{Upper frame:}} Light curve of WZ Sge in 2001, showing a rapid rise and subsequent decay punctuated by an episode of 12 echo outbursts. A freehand curve has been added to improve visibility. This article originally appeared in \citet{pat02}. Copyright 2002, Astronomical Society of the Pacific; reproduced with permission of the Editors. {\it{Lower frame:}} Results of simulations showing the occurrence of echo outbursts.}
\label{fig:trunc}
\end{figure*}

\section{Results}
Simulations were performed using the code described above. The results presented used a very high viscosity, such that the Shakura-Sunyaev alpha parameters in the hot and cold states are $0.6$ and $0.06$ respectively. The triggers are also placed rather close together and these initial calculations were performed using a low resolution of 30,000 particles, since they were part of a large parameter study. The very high viscosity has the effect of speeding up the viscous processes and hence reducing processing time. The viscosity switching triggers were activated only after the disc had been filled. A weak magnetic field of  $B \sim 1 \times 10^{5} \ {\rm{G}}$ was used. As shown in Figure 1, the main outburst in our simulation is followed by a series of echo outbursts which are of a similar form to those in the observation which is also plotted. The timescale for the echo outbursts is an order of magnitude shorter than that observed. This is as expected since physical disc alphas are thought to be of order $0.1$ and $0.01$. The echoes do not at present switch off, but continue up to the end of the run.

Figure 2 shows the state of the disc during the echo outburst phase. The upper panels show surface density and the viscous alpha in the disc. The two circles represent the 2:1 and 3:1 orbital resonances. The lower panel is a plot of the azimuthally averaged surface density as a function of radius. The central hole in the accretion disc, due to the magnetic propeller, is visible in all these plots. This latter plot also illustrates the positions of the two viscosity triggers in the simulation. Together the plots show how the echo outbursts occur in the simulations. After a main outburst, some material remains in the outer disc. Spiral waves are excited, as a result of the binary potential. The associated density enhancements, when augmented by continuing mass transfer, aid the transport of matter towards the centre where, as accretion is inhibited by the magnetic propeller, an echo outburst occurs. The echo fades when the propeller is overcome by the viscous forces and accretion onto the star resumes. Echoes do not appear if the simulation is performed without the magnetic field.

\begin{figure*}[t]
\begin{center}
\resizebox{55mm}{55.0mm}{
\mbox{
\includegraphics{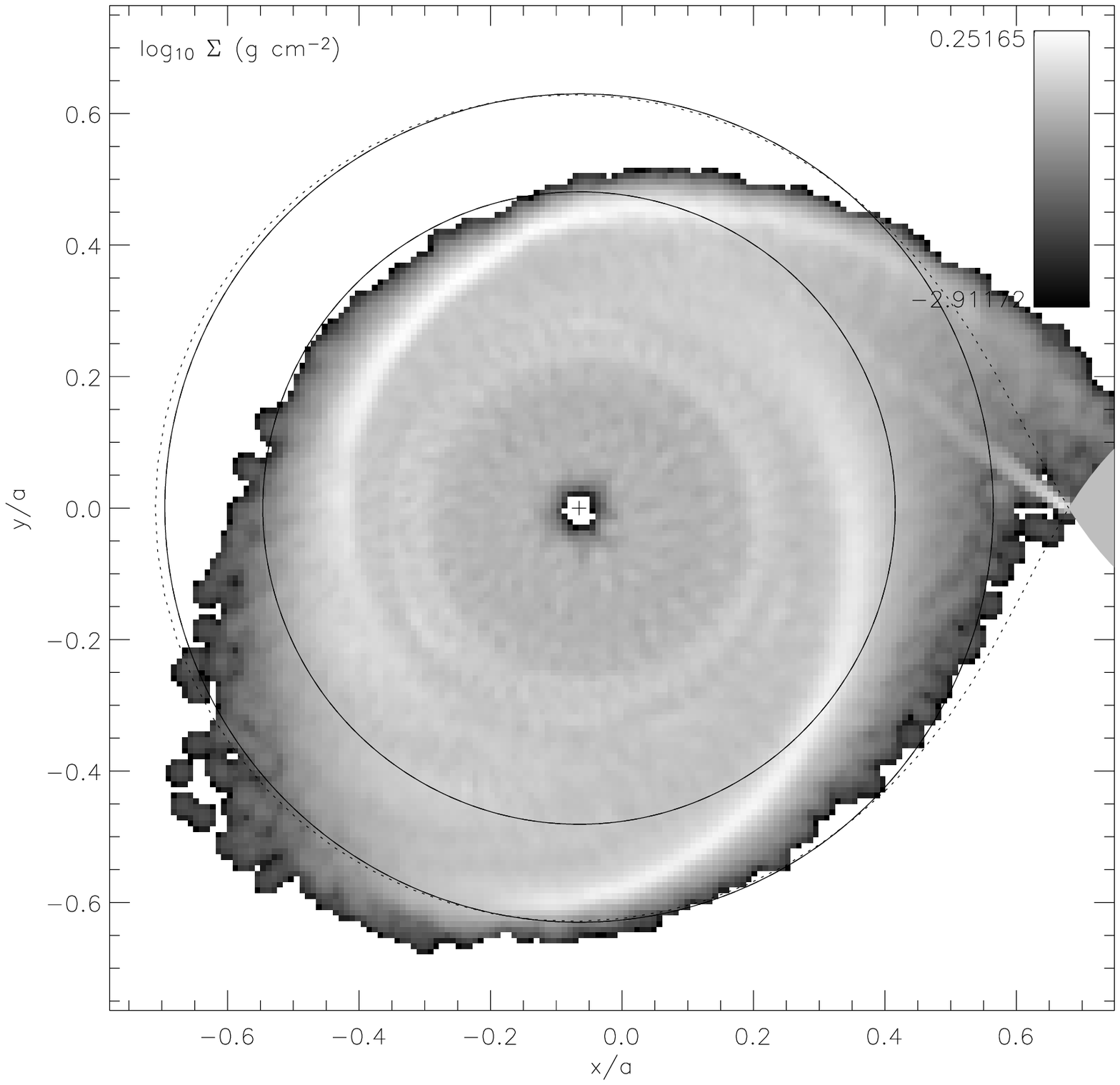}}}
\resizebox{55mm}{55.0mm}{
\mbox{
\includegraphics{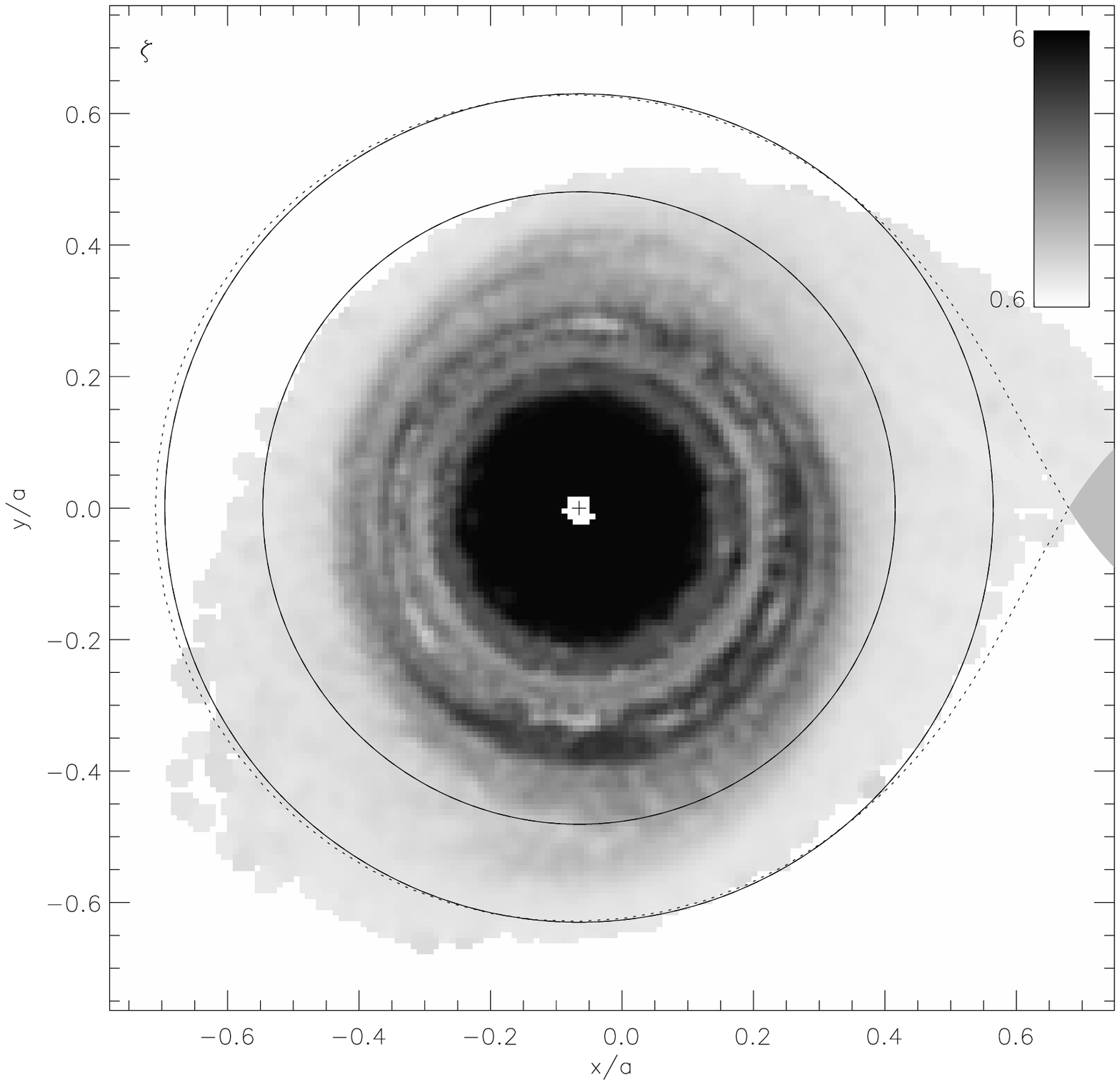}}}
\resizebox{110mm}{40.0mm}{
\mbox{
\includegraphics{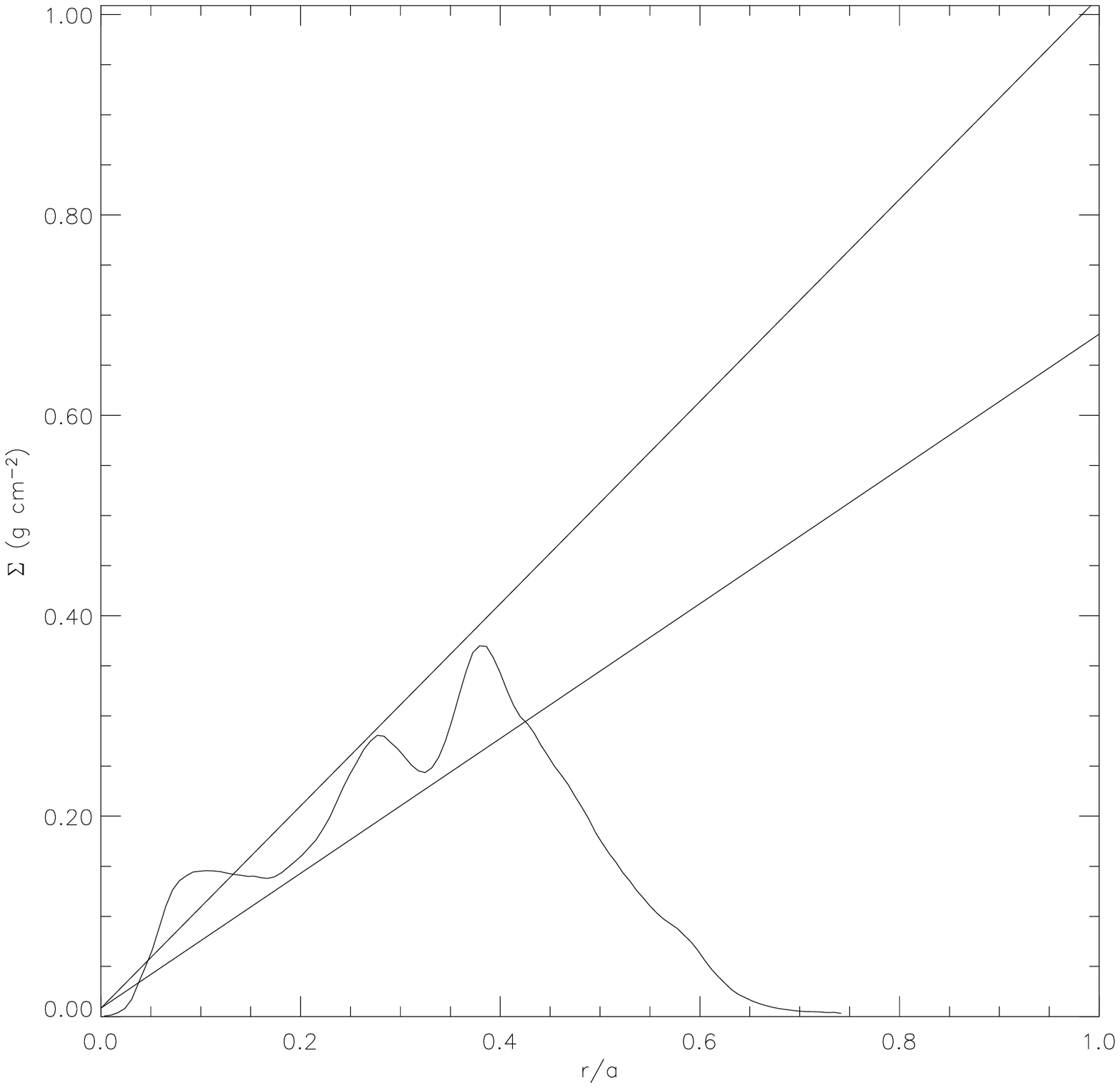}}}
\end{center}
\caption{{\it{Upper frames:}} Snapshots of the accretion disc from our simulations showing log surface density and the viscous parameter $\zeta=10 \alpha$ respectively. The binary separation is represented by $a$ in both plots. {\it{Lower frame:}} A plot of azimuthally averaged surface density as a function of radius also showing the viscosity switching triggers.}
\label{fig:sigma}
\end{figure*}

\section{Discussion}
In this paper echo outbursts are reproduced in an SPH model of WZ Sge. These echoes are similar to those observed in the light curve of the system However, it is not yet clear whether this mechanism is likely to be the one which occurs in reality. In order to establish this there are a number of refinements which will be made to the model. Firstly the viscous alphas used in the model will be brought closer to their accepted values. It is not desirable to rely on time scaling arguments in this system since neither magnetic nor resonant effects scale with alpha. The $\sim 30,000$ particles used are not sufficient to fully resolve such a disc in three dimensions, so higher resolution calculations will be made. In addition a mechanism for the termination of the echoes will be sought.

\section*{Acknowledgements}
The computations reported here were performed using the UK Astrophysical Fluids Facility (UKAFF). Research in theoretical astrophysics at the University of Leicester is supported by a PPARC rolling grant. OMM gratefully acknowledges
support through a PPARC research studentship. He would also like to thank the University of St Andrews School of Physics and Astronomy for their hospitality during a visit, when parts of this work were initiated. MRT acknowledges a PPARC
postdoctoral fellowship.

\label{lastpage}

\end{document}